\documentclass{ws-p8-50x6-00}

\begin{document}

\title{Issues of LiBeB, Oxygen and Iron Evolution}

\author{Reuven Ramaty}

\address{NASA/GSFC, Greenbelt, MD, USA, ramaty@gsfc.nasa.gov}

\author{Richard E. Lingenfelter}

\address{University of California, San Diego, USA, RICHL@cass01.ucsd.edu}  

\author{Benzion Kozlovsky}

\address{Tel Aviv University, Israel, benz@wise1.tau.ac.il}

\maketitle

\abstracts{We discuss the highlights of our recent research, 
specifically the refractory vs. volatile interpretation of the rise 
of [O/Fe] with decreasing [Fe/H], and the issue of primary vs. 
secondary evolution of Be.}

\section{Introduction}

The evolution of the light elements, Li, Be, B (LiBeB), is an 
important ingredient in studies of cosmic evolution. Closely related 
are the evolutions of O and Fe, because [LiBeB/H] are measured as 
functions of either [Fe/H] or [O/H]. In several recent papers we 
treated in detail LiBeB production\cite{ram97} by cosmic rays, and 
LiBeB, O and Fe evolution\cite{ram00a,ram00b}. We emphasized a model 
in which the cosmic ray source composition is independent of [Fe/H], 
based on cosmic ray acceleration in metal enriched 
superbubbles\cite{hig98,ling00}. In the evolutionary calculations, 
we employed a Monte Carlo approach in which supernova explosions are 
treated as discrete events in time. This approach allowed us to 
introduce delays between the time of the explosion and the 
deposition time of the nucleosynthetic products into star formation 
regions. For the LiBeB the delays are due to the transport of the 
cosmic rays. For refractory supernova synthesized elements, e.g. Fe, 
the delay is due to the slowing down time of fast moving supernova 
grains, provided that the bulk of the ejecta are in such grains, 
suggested by observation of the width of the 1.809 MeV gamma ray 
line from $^{26}$Al decay. Below we briefly summarize our most 
important results. 

\section{Discussion}

We showed that the rise of [O/Fe] with decreasing [Fe/H], recently 
confirmed\cite{isr01}, could be due to the delayed deposition of 
refractory Fe relative to the mostly volatile O. This model has the 
advantage over several other models, e.g.\cite{pran00}, because it 
provides a simple explanation for the observation that the abundance 
ratios of the refractory $\alpha$-nuclei Mg, Si, Ca and Ti relative 
to Fe do not increase with decreasing [Fe/H] below [Fe/H] $= -1$. 
The reason being that the bulk of these refractories should also be 
in grains and thus behave like Fe. On the other hand, the model 
predicts\cite{ram00b} that volatile S should behave like O, a trend 
which appears to be confirmed\cite{takeda}.

We showed\cite{ram00a} that secondary Be evolution, due to 
production by cosmic rays accelerated out of the average metal poor 
ISM in the early Galaxy, is energetically untenable, the increase of 
the O abundance at low [Fe/H] notwithstanding. On the other hand, by 
employing a cosmic ray origin model\cite{ling00}, in which the 
cosmic rays are accelerated in superbubbles out of a suprathermal 
population generated from the sputtering of, and scattering by fast 
supernova grains, we demonstrated\cite{ram00a,ram00b} that primary 
Be evolution, entirely consistent with cosmic ray energetics, can 
well account for the observations, e.g.\cite{boes99} The fact that 
the slope of log(Be/H) was found\cite{boes99} to be steeper vs. 
[O/H] than vs. [Fe/H] (1.45$\pm$0.04 and 0.96$\pm$0.04, 
respectively) appears to suggest a combination of primary and 
secondary evolutions. However, we have shown\cite{ram00b} that if 
the delay of Be deposition due to cosmic ray transport is taken into 
account, the observed trend vs. [O/H] is consistent with primary 
evolution. The often mentioned notion that the trend of Be vs. O is 
more relevant than vs. Fe is a misconception, because at low [Fe/H] 
Be is produced by cosmic ray O which is not accelerated out of the 
average ISM.

\end{document}